\documentclass{v16nufact}

\def\be{\begin{equation}}
\def\ee{\end{equation}}
\def\bea{\begin{eqnarray}}
\def\eea{\end{eqnarray}}

\newcommand{\Photo}{}

\begin{document}
\vspace*{4cm}
\title{Testable Baryogenesis in Seesaw Models\\
\small NuFact 2016}

\author{ J. Salvado$^1$, P.~Hern\'andez$^1$, M.~Kekic$^1$, J.~L\'opez-Pav\'on$^2$, J.~Racker$^1$ }

\address{$^1$Instituto de F\,isica Corpuscular (IFIC), CSIC-Universitat de
  Valencia\\
$^2$Istituto Nazionale di Fisica Nucleare (INFN) - Sezione di Genova}

\maketitle\abstracts{We revisit the production of baryon asymmetries
  in the minimal type I seesaw model with heavy Majorana singlets  in
  the GeV range. We include ``washout'' effects from
  scattering processes with gauge bosons, higgs decays and inverse
  decays, besides the dominant top scatterings. We show that in the
  minimal model with two singlets, and for an inverted light neutrino
  ordering, future measurements from SHiP and neutrinoless double beta
  decay could in principle provide sufficient information  to predict
  the matter-antimatter asymmetry in the universe. \cite{Hernandez:2016kel}}

\section{Introduction}

The origin of the observed baryon asymmetry in the Universe is one of
the unresolved fundamental questions. To dynamically generate the
this asymmetry, three Sakharov conditions need to be fulfilled:
processes that violate baryon number, processes that violate both C
and CP, and the out-of equilibrium processes should be present.  In the popular models
of baryogenesis through leptogenesis, all three conditions are
satisfied in the decay of a heavy particle, usually above TeV
scale. However, in the so-called low scale seesaw model, where the
masses of sterile neutrinos are below the electroweak scale, it is
possible to generate the asymmetry in the lepton sector via the
oscillations of the sterile states, as firstly proposed in \cite{Akhmedov:1998qx}.
Generated asymmetry is stored in the sterile sector at least until the
time of the electroweak phase transition when the sphaleron processes that
violate the baryon number become inefficient. For the asymmetry not to
be washed-out sterile neutrinos should not reach thermal
equilibrium. This reflects in the sterile neutrinos
couplings and masses, and put them to be in the $\mathcal{O}$(GeV)
scale. Note that this is exactly the mass range future experiments as
SHiP \cite{Anelli:2015pba}, DUNE \cite{Acciarri:2015uup} and FCC
\cite{Blondel:2014bra} will be sensitive to, which makes this model
potentially testable.

In this work we focus in two different questions: what are constraints
on the parameter space of the model assyming that the described
mechanism generetes all the observed baryon asymmetry, and, can we
predict the baryon asymmetry if the sterile neutrinos
are measured in SHiP or FCC together with the $\delta_{CP}$ phase in DUNE and the
neutrinoless double beta decay effective mass $m_{\beta\beta}$.

\section{Formalism}\label{subsec:prod}

The evolution of the sterile (anti)neutrino and the lepton asymmetry
in the form of chemical potential is well described by the density
matrix formalism proposed by Raffelt and Sigl \cite{Sigl:1992fn}.  The
previous analysis \cite{Asaka:2005pn,Asaka:2011wq} usually assumed the production rate of the sterile
neutrinos to be dominant by the top quark scattering and the Pauli
blocking terms in the kinetic equations are usually neglected.  Here
we derive the set of kinetic equations that beside the top quark
interactions include $2\leftrightarrow 2$ scatterings at tree level
with gauge bosons, as well as $1 \leftrightarrow 2$ scatterings
including the resummation of scatterings mediated by soft gauge
bosons.  Those effects were derived in \cite{Besak:2012qm} for the case of
vanishing leptonic potential and we re-evaluated them including the
chemical potential up to the linear order.  We also kept Fermi-Dirac
or Bose-Einstein statistic throughout and include the spectators
effects.

The evolution of sterile (anti)neutrino $r_{N/\bar{N}}\equiv \rho_{N/\bar{N}}/\rho_F$ 
 and the chemical potentials defined as $n_{B/3-L_\alpha}= {T^3 \over6} \mu_{B/3-L_\alpha}$,
 are:
{\small
\begin{eqnarray}
x H_u {d r_N\over d x} &=& -i [\langle H\rangle, r_N]  -{\langle\gamma^{(0)}_N\rangle\over 2} \{Y^\dagger Y, r_N-1\}
+\langle \gamma_N^{(1)} \rangle Y^\dagger \mu Y   -  {\langle \gamma_N^{(2)}\rangle \over 2}  \big\{Y^\dagger \mu Y,r_N\big\},\nonumber\\ 
x H_u {d r_{\bar N}\over d x} &=& -i [\langle H^*\rangle, r_{\bar N}]  -{\langle\gamma^{(0)}_N\rangle\over 2} \{Y^T Y^*, r_{\bar N}-1\} 
-\langle \gamma_N^{(1)} \rangle Y^T \mu Y^*   +  {\langle \gamma_N^{(2)}\rangle \over 2}  \big\{Y^T \mu Y^*,r_{\bar N}\big\},\nonumber\\  
x H_u {d{\mu}_{B/3-L_\alpha}\over d x} & = & {\int_k \rho_F\over
  \int_k \rho'_F}  \Bigg\{{\langle \gamma_N^{(0)}\rangle\over 2}
(Y r_N Y^\dagger- Y^* r_{\bar N} Y^T)_{\alpha\alpha} +\mu_\alpha \Bigg({\langle\gamma_N^{(2)}\rangle\over 2} (Y r_N
    Y^\dagger+Y^* r_{\bar N} Y^T)_{\alpha\alpha} \nonumber\\ 
&-&\langle\gamma_N^{(1)}\rangle {\rm Tr}[YY^\dagger I_\alpha]
\Bigg)\bigg\},\nonumber\\
\mu_{\alpha} &=& - \sum_\beta C_{\alpha\beta} \mu_{B/3-L_\beta},
\label{eq:rhonrhonbarav}
\end{eqnarray}}
{\small

where $\gamma_N$ are momentum averaged interaction rates, $Y$ is the
Yukawa matrix and $\rho_F$ is the Fermi-Dirac equilibrium distribution.

The baryon asymmetry is expressed in terms of $\mu_{B/3 -L_\alpha}$ as 
\begin{eqnarray}
Y_B \simeq 1.3 \times 10^{-3}  \sum_\alpha \mu_{B/3 -L_\alpha}.
\label{eq:yb}
\end{eqnarray}
The current value given by the Planck collaboration is,
\begin{equation}
\label{ba}
  Y_B^{\rm exp} \simeq 8.65(8) \times 10^{-11}.
\end{equation}

The numerical solution of kinetic equations is very challenging due to the
fast oscillating terms and the analytic approximations are usually
made. Even though they are useful for understanding the general
behavior of the solutions, they can not be applied to all the
parameter space.
To preform the numerical scan over all the parameter space we used
MultiNest and the solution of the equations is done numerically using 
 the public available code SQuIDS \cite{Delgado:2014kpa}.

\section{Results}

We use a Gaussian likelihood consistent with the Planck measurement
\ref{ba} and perform a Bayesian analysis of the model parameters. 
The results are summarized in
Table \ref{table}, where we have considered flat priors in all the Casas-Ibarra
parameters except the masses where we assume a flat prior in $\log_{10}\left({ M_1\over{\rm GeV}}\right)$,
within the range $M_1 \in [0.1{\rm GeV} , 10^2 {\rm GeV}]$, and two possibilities:
Blue contours,  with a flat prior in $\log_{10} \left({M_2\over GeV}\right)$ in the same range; and red contours, with a flat prior in $\log_{10} \left({|M_2-M_1|\over {\rm
GeV}}\right)$ in the range $[10^{-8}{\rm GeV} , 10^2{\rm GeV}]$.
The second option force the Bayesian sampling to explore the high
degeneracy of the mass parameters.

\begin{center}
  \begin{table}
    \begin{tabular}{|cccccccc|}
      \hline\hline
      NO & Prior & $M_1({\rm GeV})$  & $\Delta M_{12}({\rm GeV})$ & $|U_{e4}|^2$ & $|U_{\mu 4}|^2$ & $|U_{\tau 4}|^2$ & $m_{\beta\beta}({\rm eV})$ \\
      \hline
      IH & M& $-0.55^{+0.16}_{-0.38}$  & $-2.23^{+0.22}_{-0.19}$  &  $-7.2^{+0.9}_{-0.4}$&$-8.5^{+1.0}_{-0.6}$  & $-8.5^{+0.8}_{-0.7}$ & $-0.84\pm 0.55$\\
      & $\Delta$M & $0.23^{+0.68}_{-0.82}$  & $-2.36^{+0.71}_{-0.51}$  &  $-9.2^{+1.7}_{-1.4}$&$-10.1^{+1.5}_{-1.2}$  & $-9.9^{+1.4}_{-1.2}$ & $-1.48^{+0.15}_{-0.28}$\\
      \hline
      NH & M &  $-0.39^{+0.31}_{-0.42}$ & $-3.1\pm0.4$ & $-8.9^{+0.8}_{-0.7}$  & $-7.4\pm0.7$  & $-7.3^{+0.7}_{-0.5}$  & $-2.66\pm 0.20$\\
      & $\Delta$M & $0.8^{+0.82}_{-0.66}$  & $-2.76\pm 0.62$  & $-11.2^{1.4}_{-1.6}$ & $-9.9^{+1.3}_{-1.8}$  & $-10.0^{+1.3}_{-1.6}$ & $-2.62\pm 0.14$ \\
      \hline\hline
    \end{tabular}
    \caption{\label{table} Result for the ranges for the measured value eq.\ref{ba}}
  \end{table}
\end{center}

In Fig.\ref{interesting} we give the posterior probabilities of the
mass differences versus the sum of the mixing matrix elements over the
flavor, the electron mixing and the neutrinoless double beta decay
effective mass. Note that the blue region, that represent the less
fine-tuned solutions, can contribute significantly to the
$m_{\beta \beta}$ and also give the higher mixings.

\begin{figure}[t]
  \begin{center}
    \includegraphics[scale=0.3]{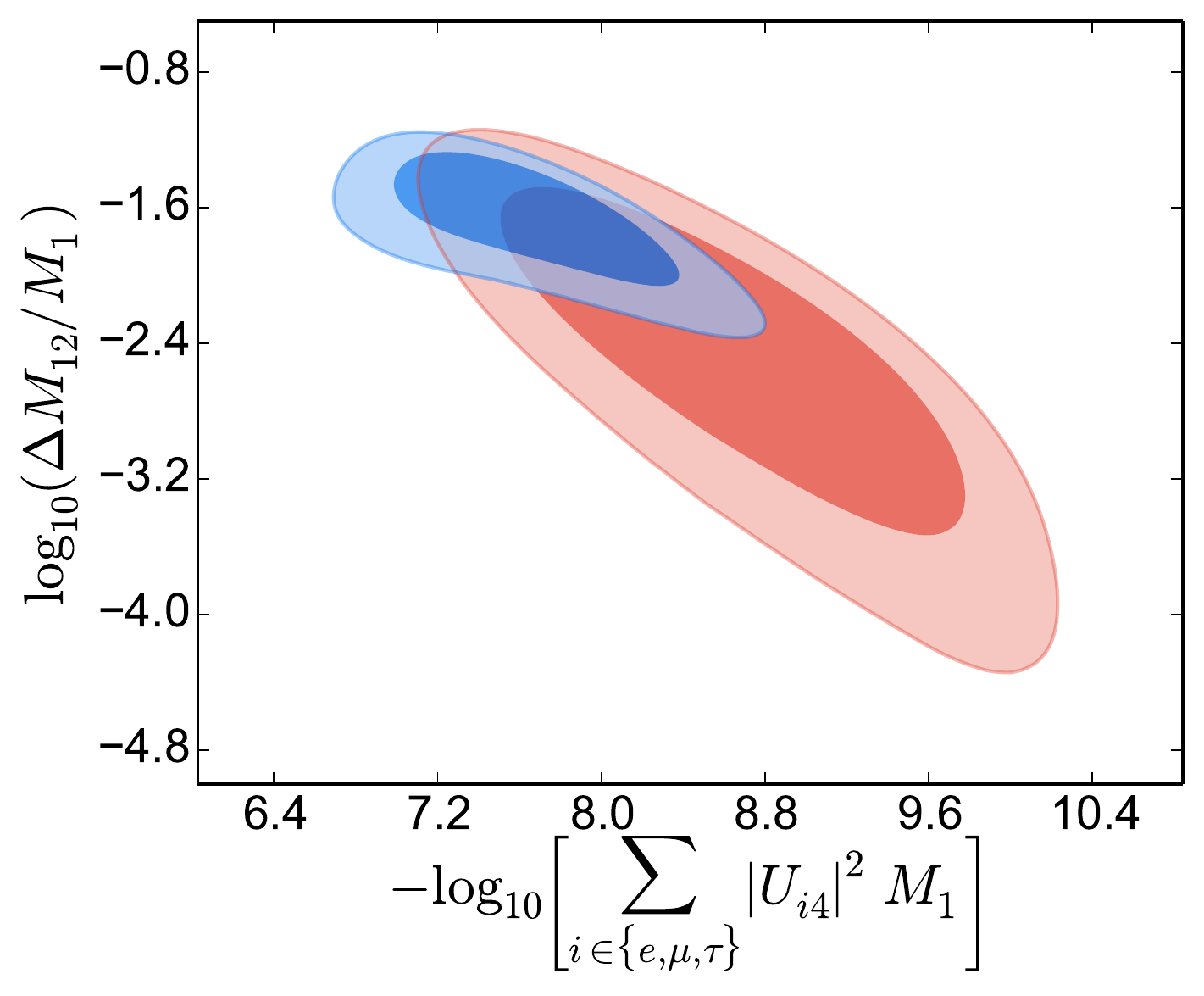} \includegraphics[scale=0.3]{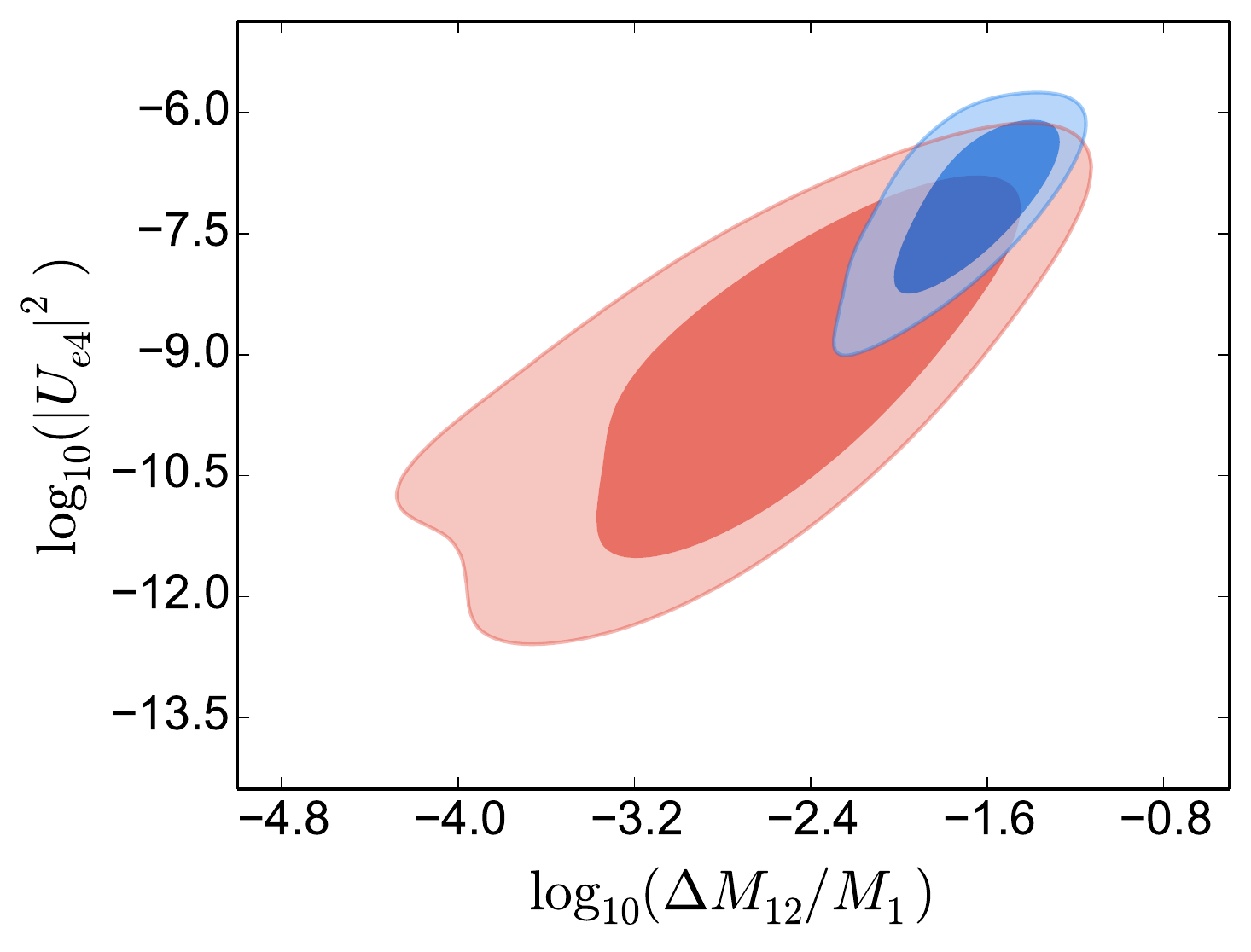}\includegraphics[scale=0.3]{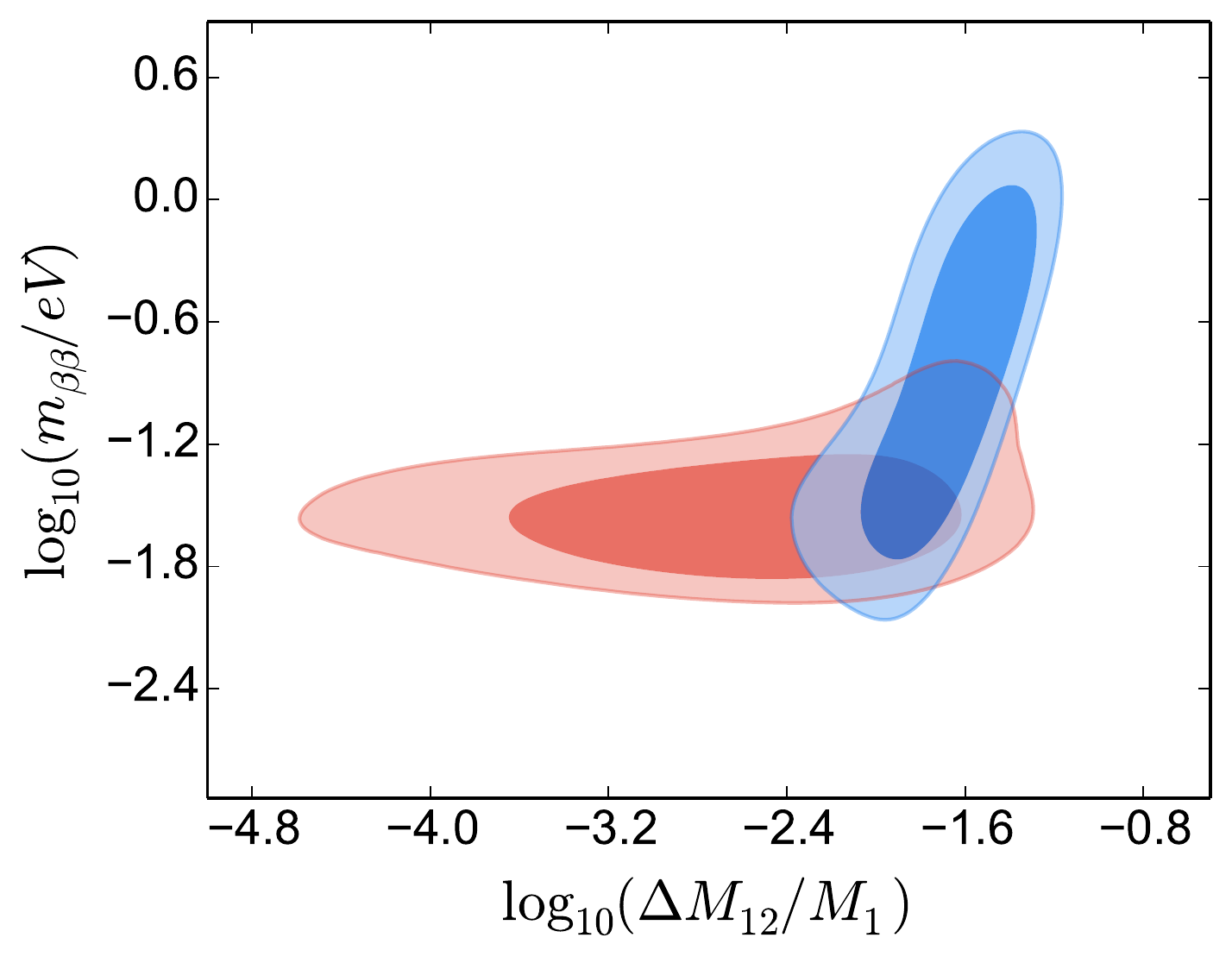}
  \end{center}
  \caption{\label{interesting} The mass degeneracy versus the sum of
    the mixing, the electron mixing and $m_{\beta \beta}$.}
\end{figure}

The mixings of the sterile state with all three neutrino flavors are
given in Fig \ref{SHIP_compare}, where the shaded region represents the
parameter space excluded from the direct searches experiment. Note
that the blue region is inside the sensitivity region of the proposed
SHiP experiment.

\begin{figure}[t]
  \begin{center}
    \includegraphics[scale=0.28]{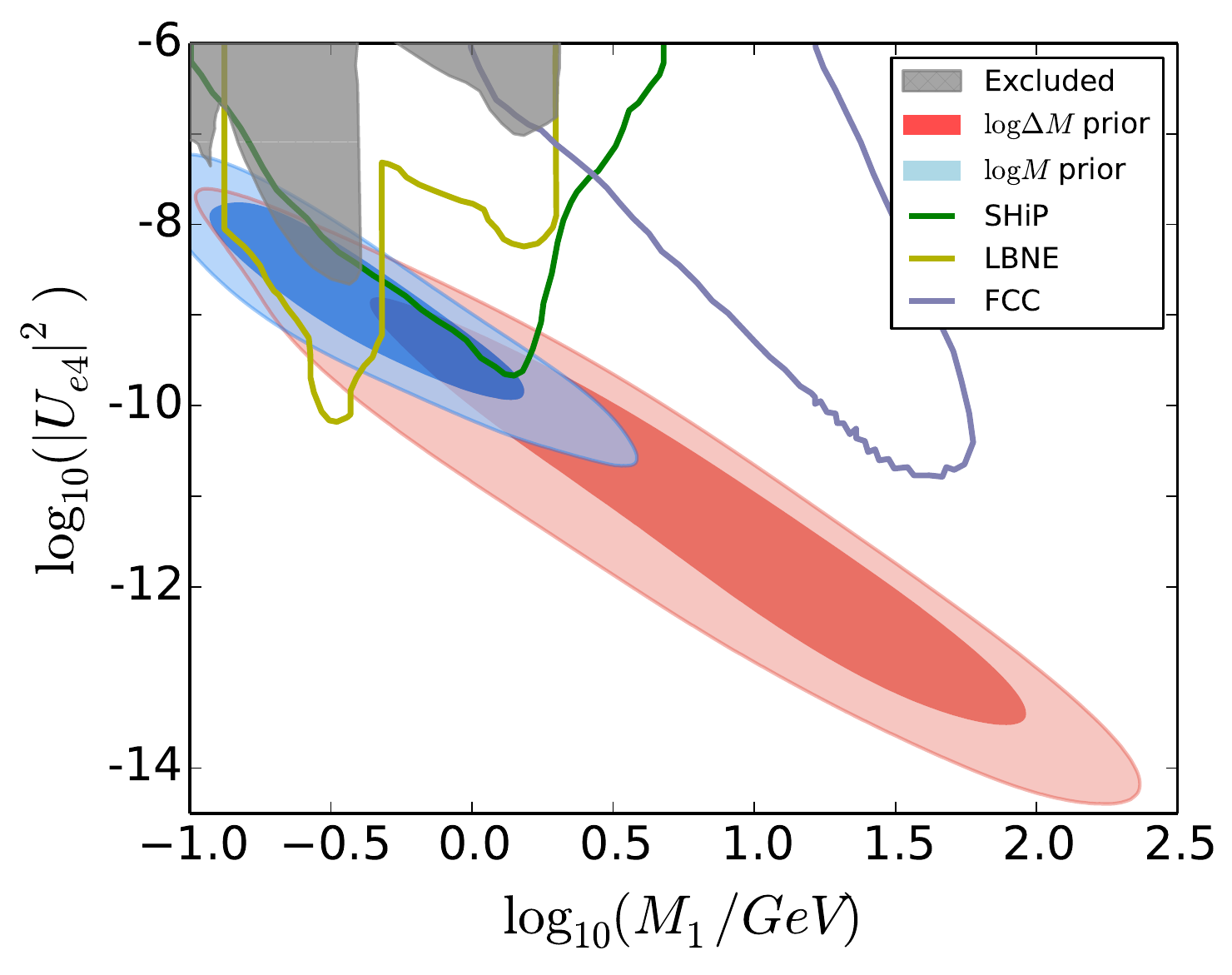} \includegraphics[scale=0.28]{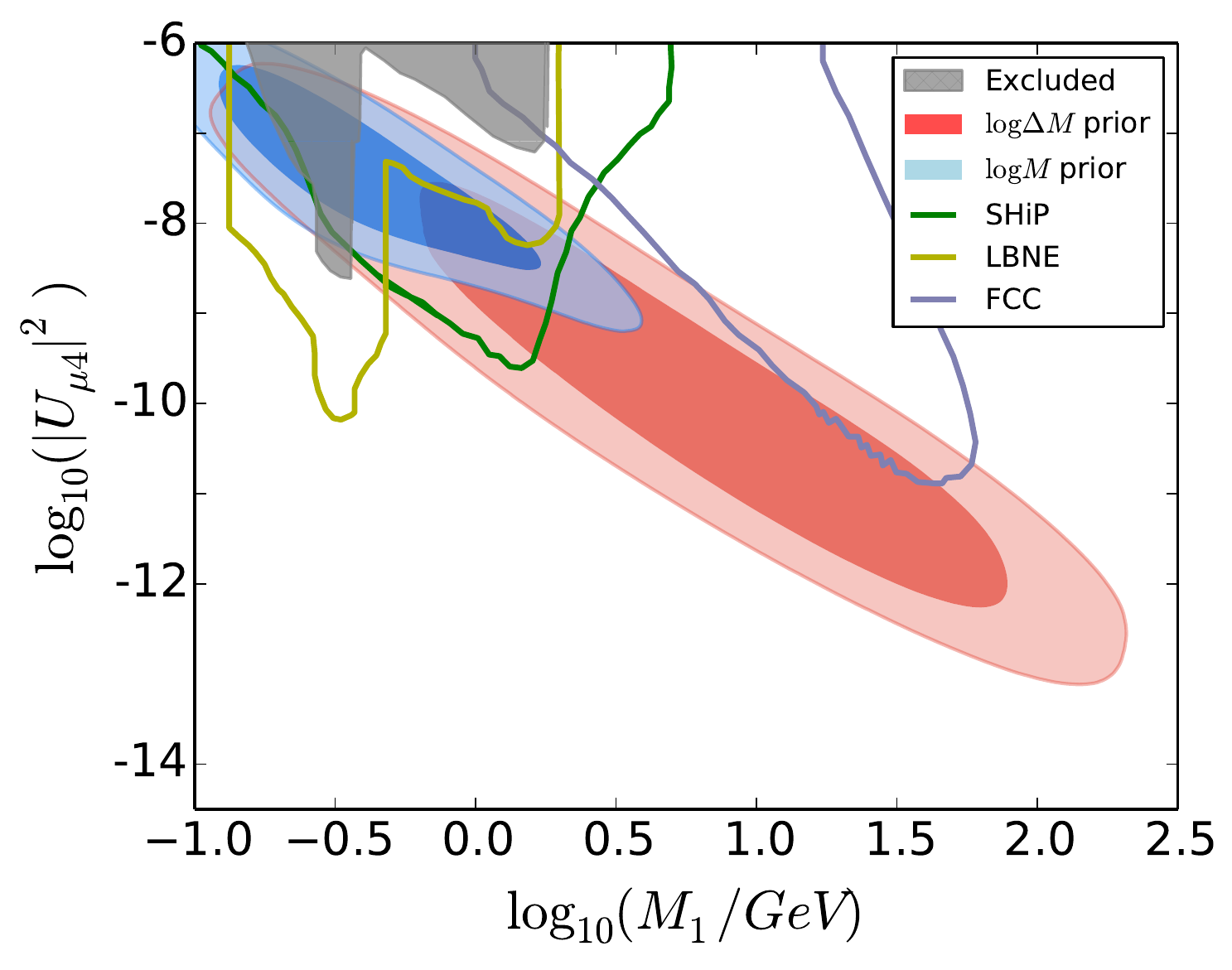}\includegraphics[scale=0.28]{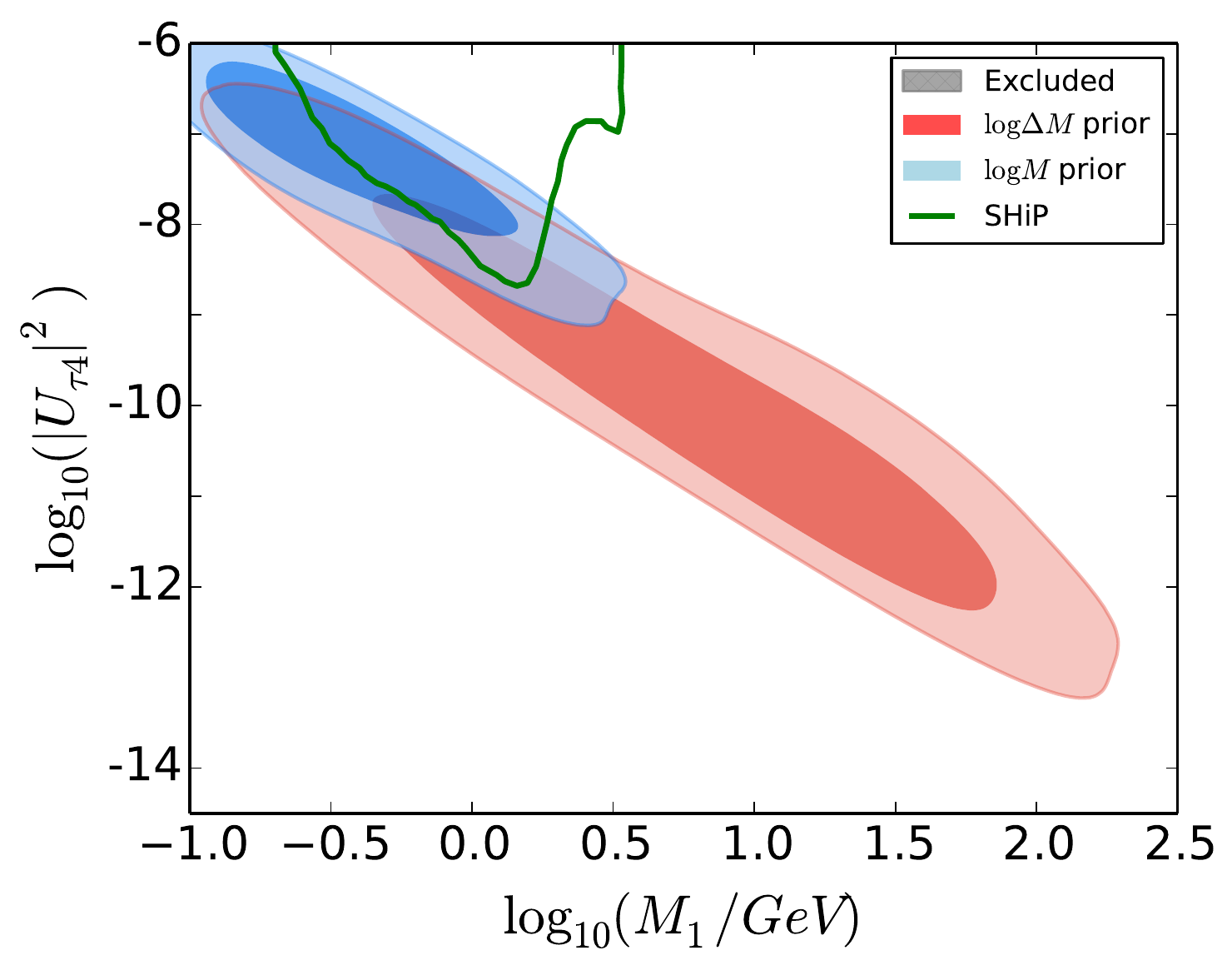}
    
    \includegraphics[scale=0.28]{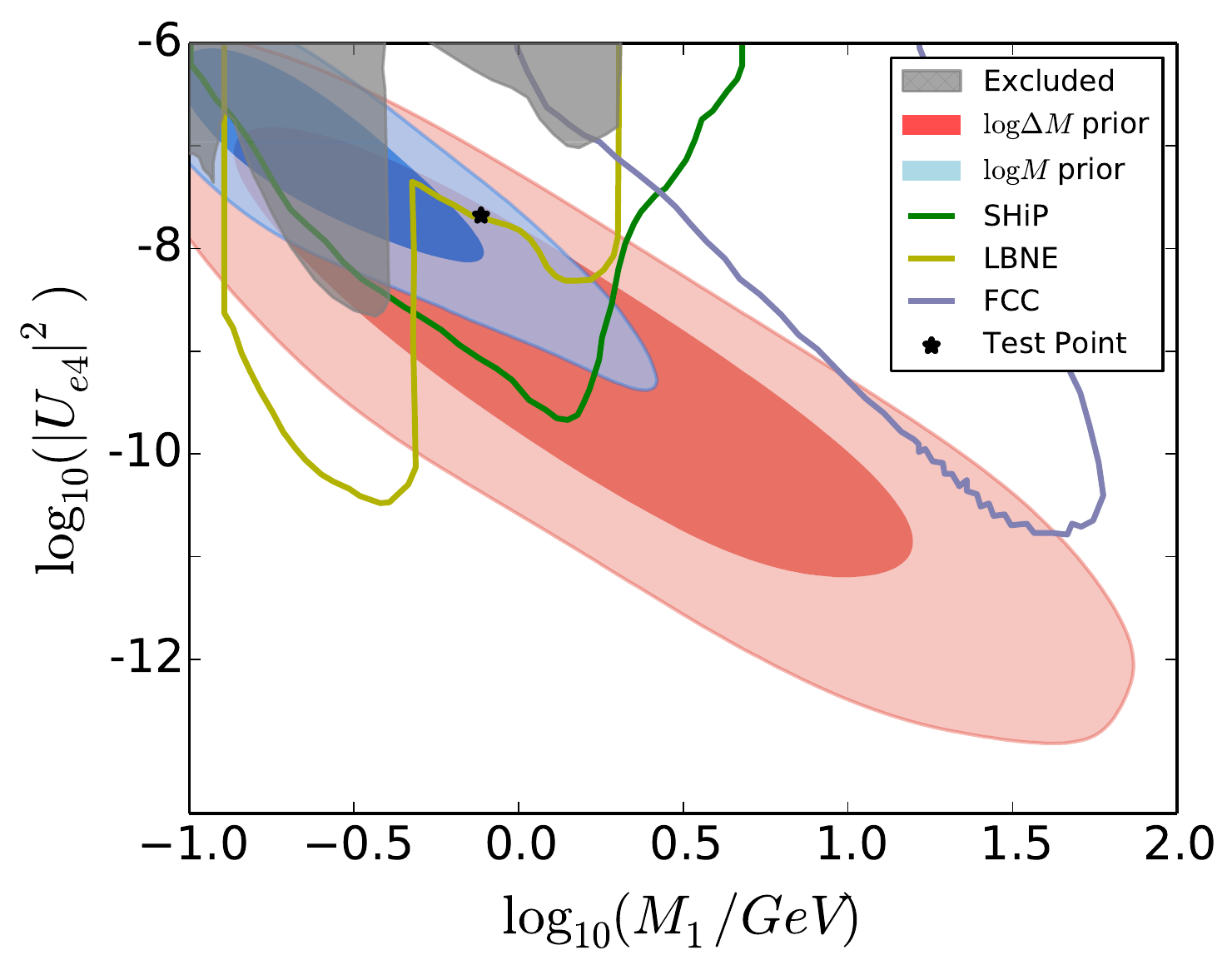}\includegraphics[scale=0.28]{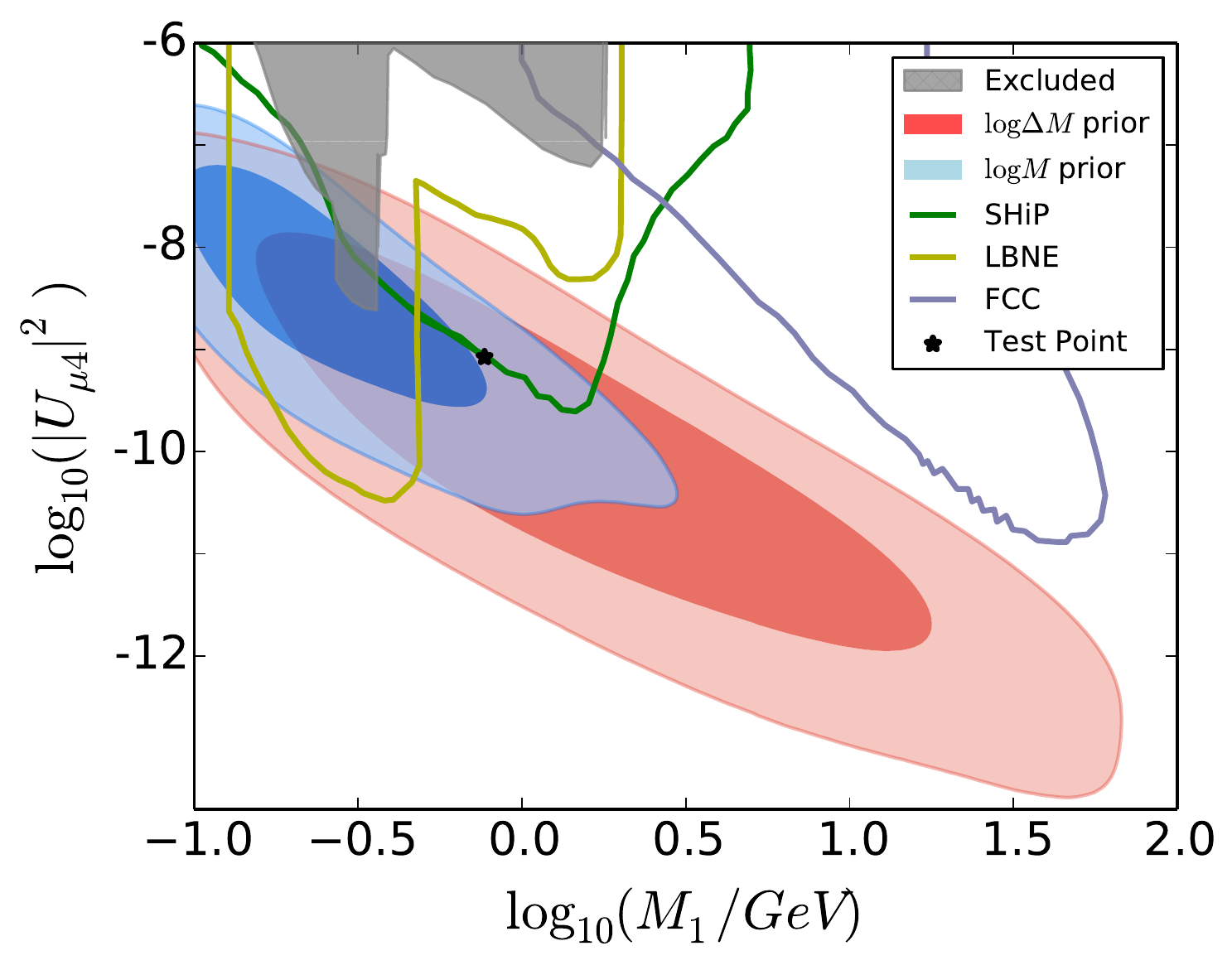} \includegraphics[scale=0.28]{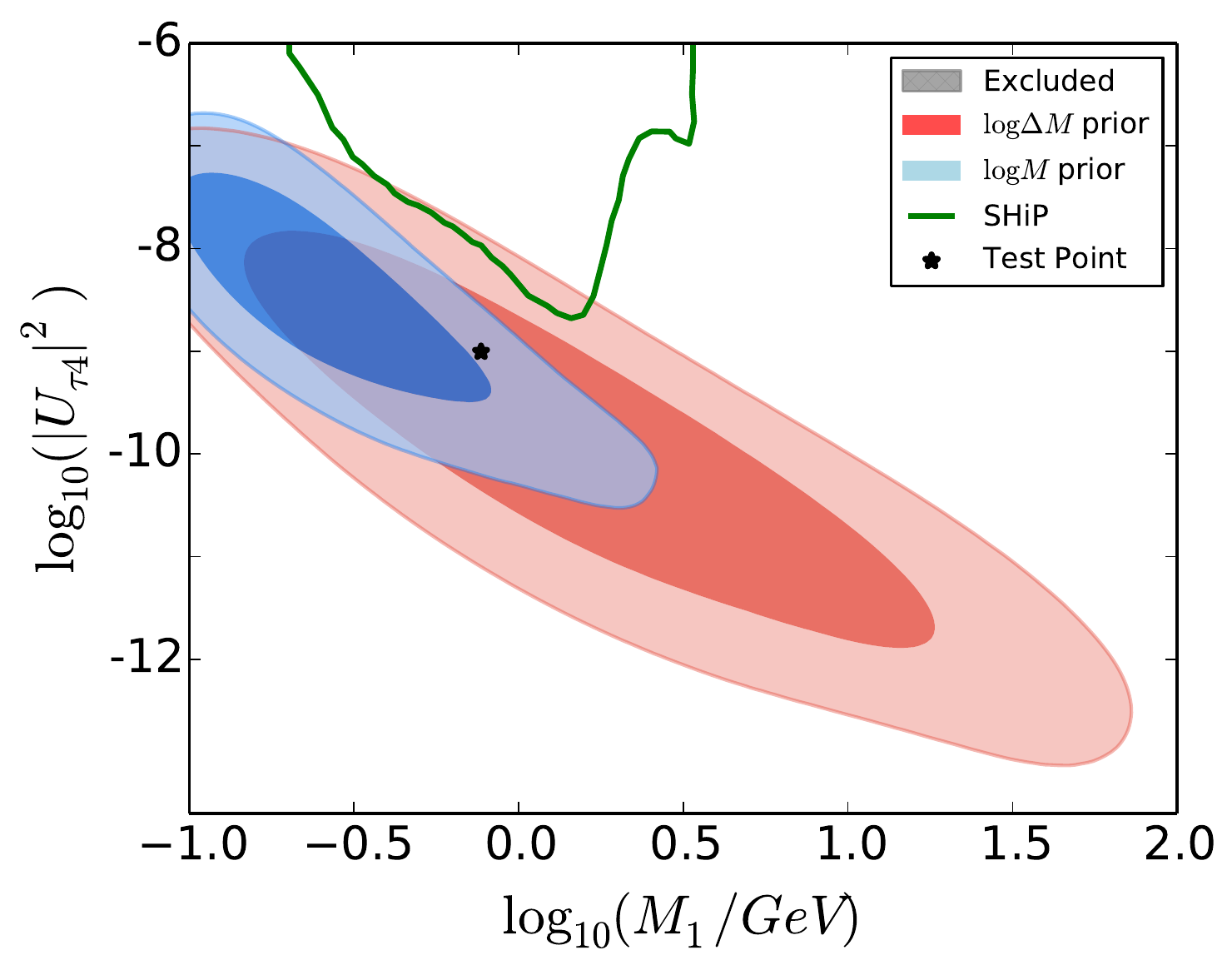} 
  \end{center}
  \caption{\label{SHIP_compare} Contours for the active sterile matrix
    elements overlapped with the expected sensitivity of the proposed
    SHiP experiment. The plots are for electron, muon, and tau from
    left to right and normal and inverted mass ordering top and bottom
  respectively.}  
\end{figure}

Next we assumed a putative measurement in SHiP, that we fix to the values
denoted by the star in the Fig \ref{SHIP_compare}.

The correlation of the
baryon asymmetry and $m_{\beta \beta}$ is presented on Fig \ref{SHIP},
where optimistic errors of 0.1\% and 1\% error in the masses  and
mixings are
considered, as well as a putative measurement of the $\delta_{CP}$ phase. 
\begin{figure}[ht]
  \begin{center}
    \includegraphics[scale=0.5]{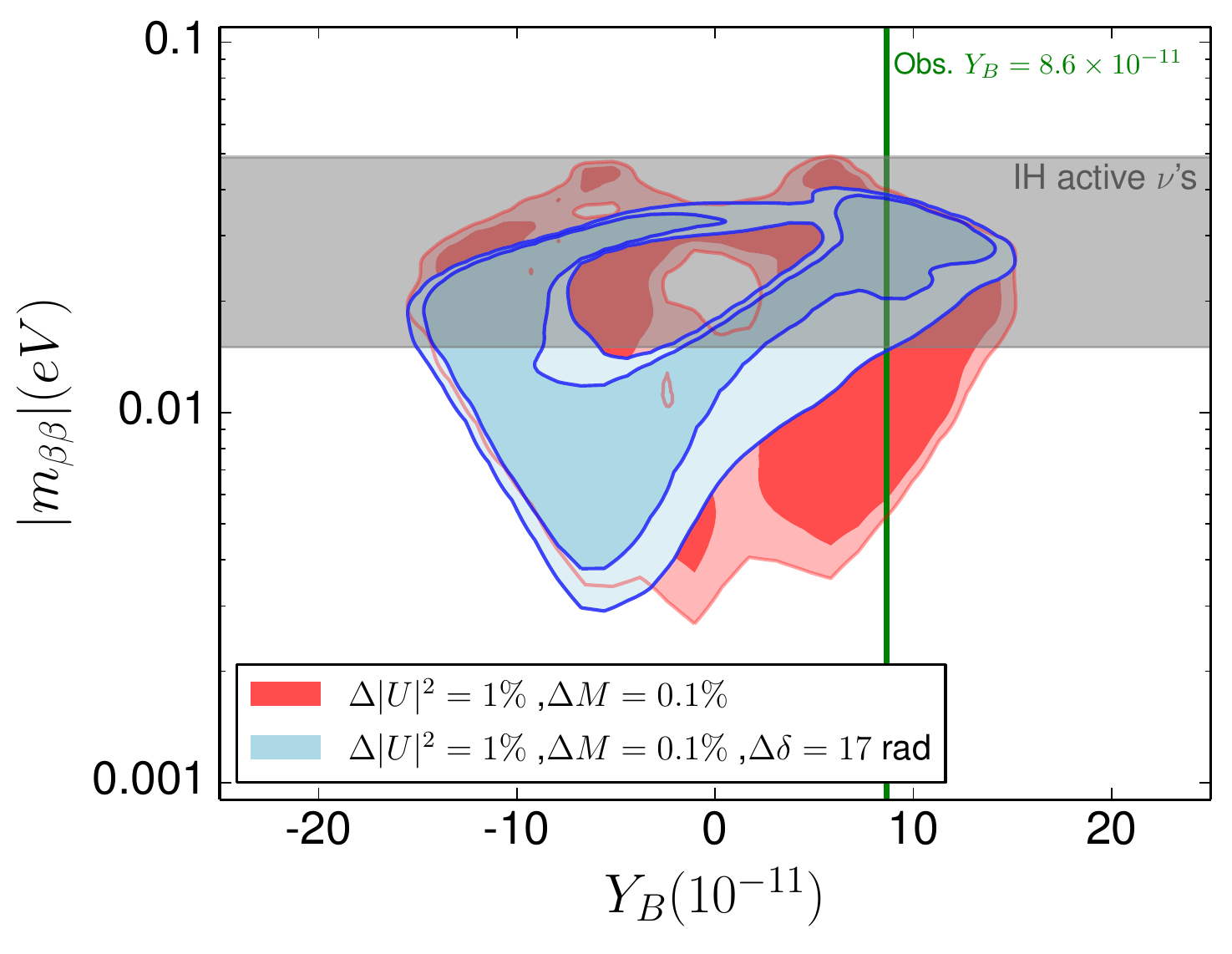}
  \end{center}
  \caption{\label{SHIP} Baryon asymmetry as a function of the
    $m_{\beta \beta}$ if a putative measure of SHiP is assumed (red region),
    and if we assume additional measurement of the $\delta_{CP}$
    phase (blue region).}
\end{figure}

The last and more detailed results about the implications of a SHiP
signal in the leptonic CP-violation can be found in the paper\cite{Caputo:2016ojx}.

\section{Conclusion}

\begin{itemize}
\item We have studied the production of the matter-antimatter
  asymmetry in low-scale ${\mathcal O}({\rm GeV})$ seesaw models. 
  
\item Our analysis include
   the washout processes from gauge interactions and Higgs
  decays and inverse decays, quantum statistics  in the computation of
  all rates, as well as spectator processes. This together with a more
  efficient numerical treatment of the equations have allowed us to
  perform the first Bayesian exploration of the full parameter space.
  
\item We have demonstrated that successful baryogenesis is possible
  with a mild heavy neutrino degeneracy, and more interestingly that
  these less fine-tuned solutions prefer smaller masses $M_i \leq
  1$GeV, which is the target region of SHiP.
  
\item If singlets with masses in the GeV range would be discovered in
  SHiP and the neutrino ordering is inverted, the possibility to
  predict the baryon asymmetry (maybe up to a sign) looks in principle viable,
  in contrast with previous beliefs.
\end{itemize}

\section*{Acknowledgments}

This work was partially supported by grants FPA2014-57816-P,
PROMETEOII/2014/050, and  the European projects
H2020-MSCA-ITN-2015//674896-ELUSIVES and H2020-MSCA-RISE-2015.


\section*{References}

\begin{thebibliography}{99}

\bibitem{Hernandez:2016kel}
  P.~Hern\'andez, M.~Kekic, J.~L\'opez-Pav\'on, J.~Racker and J.~Salvado,
  JHEP {\bf 1608}, 157 (2016)
  doi:10.1007/JHEP08(2016)157
  [arXiv:1606.06719 [hep-ph]].
\bibitem{Akhmedov:1998qx} 
  E.~K.~Akhmedov, V.~A.~Rubakov and A.~Y.~Smirnov,
  Phys.\ Rev.\ Lett.\  {\bf 81}, 1359 (1998)
  [hep-ph/9803255].

\bibitem{Anelli:2015pba}
  M.~Anelli {\it et al.} [SHiP Collaboration],
  arXiv:1504.04956 [physics.ins-det].
  
\bibitem{Acciarri:2015uup}
  R.~Acciarri {\it et al.} [DUNE Collaboration],
  arXiv:1512.06148 [physics.ins-det].

\bibitem{Blondel:2014bra}
  A.~Blondel {\it et al.} [FCC-ee study Team Collaboration],
  doi:10.1016/j.nuclphysbps.2015.09.304
  arXiv:1411.5230 [hep-ex].

\bibitem{Besak:2012qm}
  D.~Besak and D.~Bodeker,
  JCAP {\bf 1203} (2012) 029
  doi:10.1088/1475-7516/2012/03/029
  [arXiv:1202.1288 [hep-ph]].
  
\bibitem{Sigl:1992fn} 
  G.~Sigl and G.~Raffelt,
  Nucl.\ Phys.\ B {\bf 406}, 423 (1993).
  
  
\bibitem{Asaka:2011wq} 
  T.~Asaka, S.~Eijima and H.~Ishida,
  JCAP {\bf 1202}, 021 (2012)
  [arXiv:1112.5565 [hep-ph]].

\bibitem{Asaka:2005pn} 
  T.~Asaka and M.~Shaposhnikov,
  Phys.\ Lett.\ B {\bf 620}, 17 (2005)
  [hep-ph/0505013].

\bibitem{Delgado:2014kpa}
  C.~A.~Argüelles Delgado, J.~Salvado and C.~N.~Weaver,
  Comput.\ Phys.\ Commun.\  {\bf 196} (2015) 569
  doi:10.1016/j.cpc.2015.06.022
  [arXiv:1412.3832 [hep-ph]].

\bibitem{Caputo:2016ojx} 
  A.~Caputo, P.~Hernandez, M.~Kekic, J.~López-Pavón and J.~Salvado,
  arXiv:1611.05000 [hep-ph].


\end{thebibliography}
\end{document}